# Narrow Linewidth Laser Based on Extended Topological Interface States in One-Dimensional Photonic Crystals

## Author Information

Xiao Sun[1*], Zhibo Li[1], Yiming Sun[1], Yupei Wang[1], Jue Wang[1], Huihua Cheng[1], Cong Fu[1], John H. Marsh[1], Anthony E. Kelly[1], Lianping Hou[1*]

1. James Watt School of Engineering, University of Glasgow, Glasgow, G12 8QQ, U.K.

* Correspondence: Xiao Sun (Xiao.Sun@glasgow.ac.uk)

James Watt School of Engineering University of Glasgow

Glasgow, G12 8QQ, U.K

Email: Xiao.Sun@glasgow.ac.uk

* Correspondence: Lianping Hou (Lianping Hou @glasgow.ac.uk)

James Watt School of Engineering University of Glasgow

Glasgow, G12 8QQ, U.K

Email: Lianping.Hou@glasgow.ac.uk

## Abstract

Recent advances in topological one-dimensional photonic crystal concepts have enabled the development of robust light-emitting devices by incorporating a topological interface state (TIS) at the cavity center. In this study, we theoretically and experimentally demonstrate a one-dimensional TIS-extended photonic crystal (1D-TISE-PC) structure. By integrating a linearly dispersive zero-index one-dimensional photonic crystal structure with a four-phase shift sampled grating, photons propagate along the cavity without phase differences, enhancing the robustness to material variations and extending the TIS. Our findings indicate that extending the TIS promotes a more uniform photon distribution along the laser cavity and mitigates the spatial hole burning (SHB) effect. We fabricated and characterized a 1550 nm sidewall 1D-TISE-PC semiconductor laser, achieving stable single-mode operation across a wide current range from 60 to 420 mA, with a side-mode suppression ratio of 50 dB. The 1D-TISE-PC structure exhibited a linewidth narrowing effect to approximately 150 kHz Lorentzian linewidth. Utilizing reconstruction equivalent-chirp technology for the 4PS sampled grating enabled precise wavelength control in 1D-TISE-PC laser arrays, achieving a wavelength spacing of 0.796 nm ± 0.003 nm. We show that the TIS still exists in the TISE cavity and topological protection is preserved. Its mode extension characteristics mitigate the SHB so narrow the linewidth. We argue that the design simplicity and improvement of the fabrication tolerance make this architecture suitable for high-power and narrow-linewidth semiconductor lasers development.

## Introduction

Topological photonic crystals have emerged as a revolutionary concept in the field of photonics, offering unique advantages for the manipulation and control of light. These advanced structures leverage the principles of topological invariants of the photon wavefunction in the optical dispersion band, to create robust and versatile optical cavities[1,2], filters[3], modulators[4] as well as nonlinear and quantum devices[5,6]. Topological microcavity lasers designed based on the bulk band topologies have been demonstrated[7-13]. The cavity design of these lasers relies on the excitation of topological boundary states at the interface of photonic structures in different topological Zak phases. This topological interface state (TIS) exhibited single-mode lasing and a high-quality factor[14-16].

Conventional distributed feedback Bragg (DFB) semiconductor products incorporate key grating elements that are mathematically equivalent to one-dimensional (1D) topological photonic crystal models[17,18]. These models include states such as Jackiw–Rebbi zero mode[19], Su–Schrieffer–Heeger (SSH) edge state[20], and Schawlow-Townes two-mirror strategy [21]. The π-phase-shifted Bragg grating forms a TIS within the band gap, characterized by distinct Zak phases, allowing a single mid-gap mode to lase at the Bragg frequency where the grating feedback is strongest.

However, the traditional design of π-phase-shifted DFB laser cavities faces significant challenges. According to the Schawlow-Townes two-mirror strategy[22], the fundamental optical mode has a high intensity at the center of the π-phase-shifted DFB laser cavity and low intensity at the outer boundaries of the device. This distribution depletes the optical gain at the center of the cavity but not at the edges, leading to a spatial hole burning (SHB) effect. SHB diminishes output power, generates multimode lasing, lowers the side-mode suppression ratio (SMSR), and establishes a linewidth floor that contributes to linewidth broadening [23]. As a result, achieving high power and narrow single longitudinal mode(SLM) output by simply increasing the length of the laser cavity is challenging for π-phase-shifted DFB lasers.SHB diminishes the output power, generates multimode lasing, lowers the side-mode suppression ratio (SMSR), and establishes a linewidth floor that contributes to linewidth broadening, and may lead to multimode lasing [23]. As a result, achieving high power with a narrow single longitudinal mode (SLM) output by simply increasing the length of the laser cavity is challenging for π-phase-shifted DFB lasers.

In this paper, we explore the integration of topological photonic crystal concepts into the design of advanced laser systems. We demonstrate a one-dimensional TIS-extended photonic crystal (1D-TISE-PC) grating structure. This structure utilizes a laterally modulated four-phase-shifted (4PS) sampled 1D photonic crystal grating to create a Dirac-like cone at 1550 nm, resulting in a zero-index waveguide. This waveguide connects two non-modulated 4PS sampled 1D photonic crystal gratings with distinct Zak phases. The photons can propagate along the 1D-TISE-PC waveguide without a phase difference, making the system robust to material changes and maintaining SLM operation. The fundamental optical mode in the center of the cavity is efficiently extended to the edge thus the photon density along the cavity is uniform, reducing the depletion of optical gain at the central area.

We fabricate a sidewall 1D-TISE-PC semiconductor laser at 1550 nm wavelength. Our experimental results demonstrate that the 1D-TISE-PC structure suppresses the SHB effect and reduces the linewidth to 150 kHz, compared to traditional short π-phase shift segments with the linewidth larger than 1 MHz. The 1D-TISE-PC structure can also suppress the ±1 trivial Tamm modes, achieving pure SLM operation over a wide range of operating currents from 60 mA to 420 mA, even without anti-reflection (AR) facet coatings, and with an SMSR of up to 50 dB.

Building upon its 4PS sampled structure, we also successfully achieved a high-precision wavelength control in 1D-TISE-PC laser by reconstruction equivalent-chirp (REC) technology[24]. An eight-channel 1D-TISE-PC laser array is demonstrated with a 0.796 nm channel wavelength spacing with a linewidth lower than 300 kHz in each channel. Building upon the 4PS sampled structure, we also successfully demonstrated high-precision wavelength control in a 1D-TISE-PC laser by reconstruction equivalent-chirp (REC) technology[24]. An eight channel 1D-TISE-PC laser array is demonstrated with a 0.796 ± 0.003 nm channel wavelength spacing with a linewidth lower than 300 kHz in each channel.

## Results

### Design principle of the 1D-TISE-PC structure

The traditional π-phase shifted uniform Bragg grating is shown in Fig. 1a. This structure consists of two Bragg reflection mirrors with a π-phase shift between them. Figure 1b presents the structure of the 1D-TISE-PC, which includes left and right reflection grating mirrors ($L_{LG}$ and $L_{RG}$) with inversion symmetry, and a central topological interface state extending (TISE) cavity ($L_M$). Here, $\Lambda_M$ is the lateral modulation period and $\Lambda$ denotes the cell period. There are two forms of cell: a uniform Bragg grating structure ($\Lambda = \Lambda_B$) and a sample grating structure ($\Lambda = \Lambda_S$) and their modulation process in the TISE cavity is shown in Fig. 1c. and Fig. 1d respectively, where $\Lambda_B$ is the Bragg period and $\Lambda_S$ is the sampling period, $n_1$ and $n_2$ refer to two different refractive index segments. In uniform grating structures the positions of the $n_1$ and $n_2$ are changed with a shift step $\Delta\Lambda$ while maintaining a constant $\Lambda_B$ during each modulation cycle. Conversely, in the sample grating structure, the changes occur in the positions of the sampling and non-sampling segments, with $\Lambda_S$ kept constant. It can be mathematically demonstrated that the lateral modulation in the uniform grating system is equivalent to that in the sampling grating system (see supplementary Part A). Within the modulation period $\Lambda_m$, an alignment factor, $\Delta\Lambda$ quantifies the adjustment of the phase shift step across each sampling period $\Lambda_m$. After a lateral modulation period $\Lambda_m$, the cell returns to the initial state, resulting in $\Delta\Lambda = \Lambda$. For a lateral modulation period $\Lambda_m$ with $n^{th}$ modulation, it holds that $\Lambda_m = n\Lambda$ and $\Delta\Lambda = \Lambda/n$.

It has been noted that the grating elements in uniform or sampled gratings are mathematically equivalent to the supercell in one-dimensional photonic crystals (1D-PCs) in the SSH model. Therefore, these structures can be analyzed within the framework of a 1D topological system. Figure 1e illustrates, from left to right, the calculated 1D band structures of the left grating, the TISE cavity, and the right grating. Both the left and right gratings exhibit two inversion centers, while the energy band in the TISE cavity is degenerate. For the *j*-th energy band of the topological photonic crystal (TPC), the Zak phase is defined as follows[25,26]:

$$\theta_j^{Zak} = \int_{-\pi/\Lambda}^{\pi/\Lambda} \left[ i \int_{unitcell} dz \varepsilon(z) u_{j,k}^*(z) \partial_k u_{j,k}(z) \right] dk \quad (1)$$

Where $u_{j,k}(z)$ is the periodic-in-cell part of the Bloch electric field eigenfunction of a state on the *j*th band with wave vector *k* and $\varepsilon(z)$ is the dielectric function. The Zak phase can be determined by evaluating the electric field distribution within the unit cell. The general solution of the electric field $E(z)$ can be given as

$$E(z) = R(z)\exp(-in(z)k_0) + S(z)\exp(in(z)k_0) \quad (2)$$

$R(z)$ and $S(z)$ represent the forward and backward propagating electric fields, respectively, and $k_0$ is the wave vector. The electric field can be found analytically using the transfer-matrix method (see Supplementary Part B). The Zak phase is quantized at either 0 or $\pi$ when the origin is chosen to be one of the inversion centers. The Zak phase of each isolated band is calculated using Eq. (1) and is marked in Fig. 1e. Here, we only consider the bandgap near the lasing wavelength of 1550 nm and ignore other gaps. Due to the distinct Zak phase, a TIS exists in these photonic gaps when combining the left and right gratings. The band-degenerated TISE cavity provides topological protection of the phase transition from the left to the right gratings, thereby expanding the TIS mode.

The length of shift step $\Delta \Lambda$ also influences the degeneracy of the light band in the TISE cavity. Consider a constant cell period $\Lambda$ in a TISE cavity: for *n* = 5 ($\Lambda_m = 5\Lambda$, $\Delta \Lambda = \Lambda/5$) band dispersion is still present at the center wavelength, and thus the TISE cavity cannot achieve a topological-protected transition, as illustrated in Fig. 1f. For *n* = 20 ($\Lambda_m = 20\Lambda$, $\Delta \Lambda = \Lambda/20$), the band becomes degenerate, as shown in Fig. 1g. This result indicates that the shift step $\Delta \Lambda$ must be sufficiently small. Moreover, the sampled grating offers better fabrication tolerances than the uniform grating. The period of the sample grating can reach several micrometers, whereas the first-order Bragg grating period of the uniform grating is only around 243 nm at a wavelength of 1550 nm. The precision of the TISE cavity length can be enhanced by a factor of $(\Lambda_S/\Lambda_0 +1)^2$ compared with a uniform grating, typically exceeding two orders of magnitude[27].

Figure 2 illustrates the electrical field distribution in the 1D-TISE-PC cavity. The normalized electric field distribution is investigated using both the full-wave finite element method (FEM) and coupling-wave theory numerical analysis (see Supplementary Part B). The numerical analysis aligns well with the finite-difference FEM results. The total length of the 1D-TISE-PC cavity ($L = L_{LG} + L_M + L_{RG}$) is set at 50 μm, with $L_M$ set at 0, 8.5, and 15 μm to minimize memory resource requirements in the FEM. The refractive index $n_1$ and $n_2$ are set at 3.165 and 3.215 respectively.

It is found that with longer $L_M$, the $0^{th}$ TIS mode is expanded in the TISE cavity, improving the field uniformity and enabling single-mode operation. The two peaks of the $\pm 1^{st}$ trivial Tamm states[28] side mode electrical field shift towards the cavity center. The calculated transmission spectra of the 1D-TISE-PC as a function of the TISE cavity length $L_M$ are shown in Fig. 2(d). We found that there is an optimal $L_M$ that allows the TIS to suppress the Tamm states, achieving the highest SMSR. Notably, for $L_M = 8.5$ μm, the SMSR between the $0^{th}$ TIS mode and the $\pm 1^{st}$ side modes reaches its maximum value, which is better than that for $L_M = 0$ μm (0.615) or $L_M = 15$ μm (0.486), demonstrating effective apodization as illustrated in Fig. 2(e).

It is also noted that inserting the TISE cavity affects the coupling coefficient κ compared to the π-phase shift only ($L_M = 0$ μm). The equivalent $κ$ of 1D-TISE-PC is given as (see supplementary Part C):

$$\kappa = \kappa_0 \left(1 - \frac{L_M}{L}\right) \quad (3)$$

Where $\kappa_0$ is the coupling coefficient for $L_M = 0$ μm. An oversized $L_M$ will reduce the TIS mode coupling and decrease the SMSR or lead to multi-longitudinal modes. This result indicates that the TISE cavity length should be adjusted appropriately to increase the electrical field uniformity and apodize the $\pm 1^{st}$ side modes.

**Fig. 1: Design principles of 1D-TISE-PC structure**

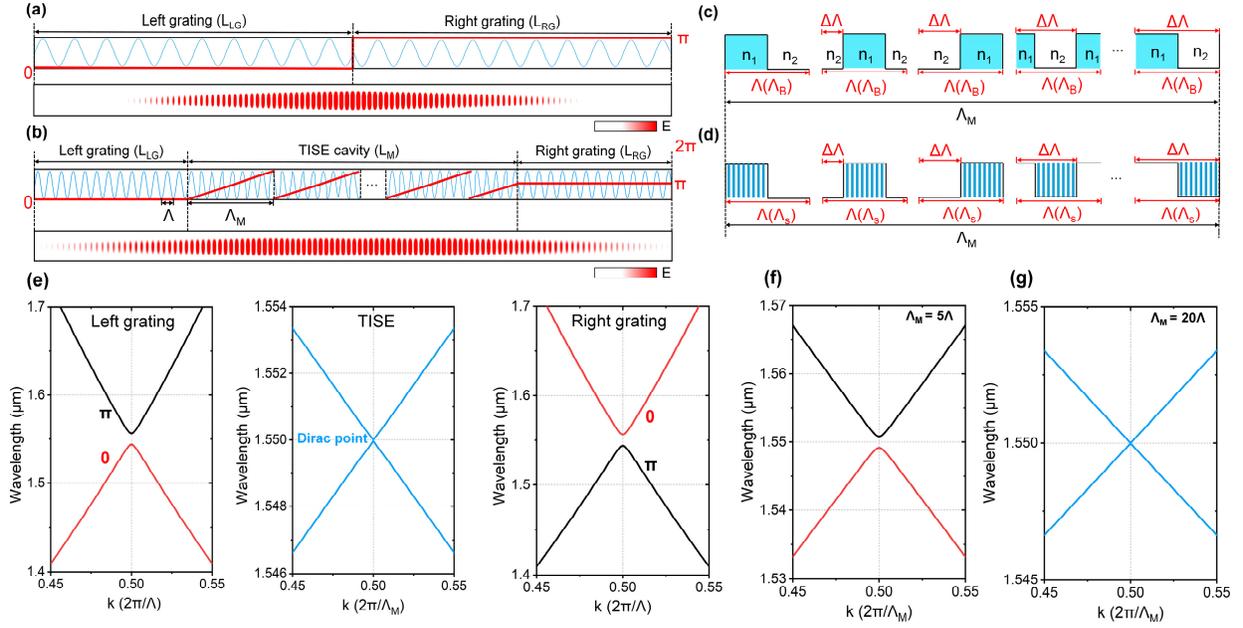

Structure and phase transfer of **a** traditional π-phase-shifted grating structure and **b** 1D-TISE-PC structure. The red line shows the phase variation in each period. The lateral modulation of the **c** Bragg uniform grating and **d** sampled grating in one modulation period $\Lambda_m$. **e** The calculated optical band structures of the left reflection grating ($L_{LG}$), topological interface states extending (TISE) grating ($L_M$), and right reflection grating ($L_{RG}$). The left-side and right-side gratings have distinct Zak phases and the TISE range to form a Dirac-like cone dispersion at $k = 0.5$. **f** and **g** Optical band structures of the TISE grating when $\Lambda_m = 5\Lambda$ and $\Lambda_m = 20\Lambda$.

**Fig. 2: Electrical field distribution in 1D-TISE-PC structure**

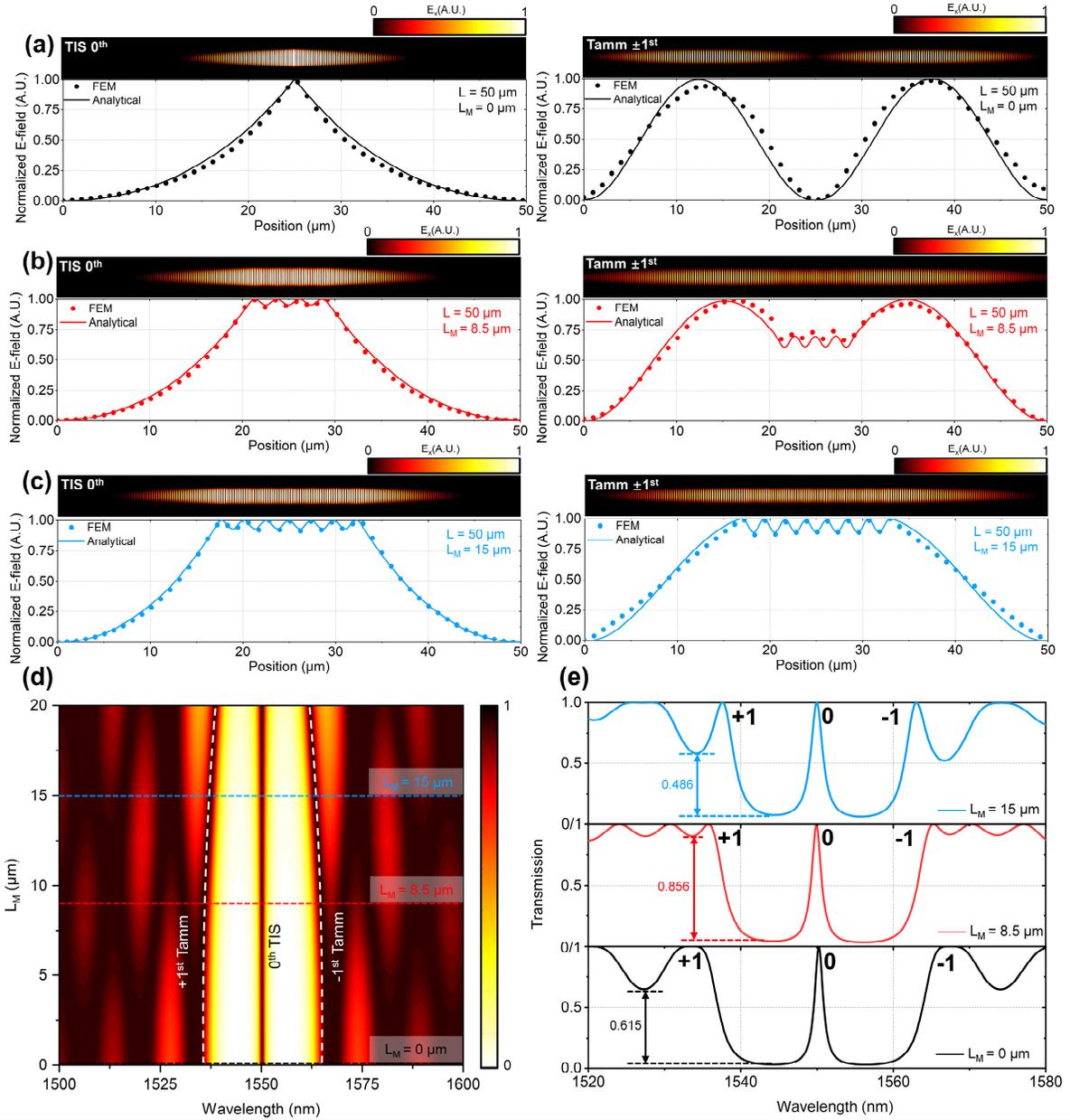

Normalized electric field distributions and transmission spectrum of the $0^{th}$ TIS mode and $\pm 1^{st}$ trivial Tamm states side modes of the 1D-TISE-PC cavity, the total length of the cavity ($L = L_{LG} + L_M + L_{RG}$) is 50 μm. **a** $L_M = 0$ μm, which is non-expanded TIS mode. **b** $L_M = 8.5$ μm. **c** $L_M = 15$ μm. The normalized electric field extracted from field profiles of the FEM simulation (dots) agrees with the numeral analysis. **d** Calculated transmission spectra of the 1D-TISE-PC as a function of $L_M$. **e** transmission spectra for $L_M = 0$ μm, $L_M =$ 8.5 μm and $L_M = 15$ μm, respectively.

## Fabrication and characterization of 1D-TISE-PC laser

The epitaxial structure used for the devices was based on the AlGaInAs/InP material system and contains five quantum wells (QWs) and six barriers [29]. The epitaxial structure used for the devices was based on the AlGaInAs/InP material system and contained five quantum wells (QWs) and six quantum barriers with a QW confinement factor of 5%. The room temperature photoluminescence (PL) wavelength of the QWs was designed to be at 1530 nm. The grating ridge waveguide width was designed to be 2.5 μm, with a sidewall corrugation depth of 250 nm. The ridge height was 1.92 μm, as illustrated in Fig. 3a. The 1D-TISE-PC structure was designed as a third-order four-phase-shift (4PS) sampled grating, with the center wavelength calculated as (see Supplementary Part D):

$$\lambda = (n_1 + n_2) \frac{\Lambda_0 \Lambda_s}{3\Lambda_s - \Lambda_0} \quad (4)$$

Here we take the effective index $n_1 = 3.1967$ and $n_2 = 3.1935$, the seed grating period $\Lambda_0 = 694$ nm, and center wavelength $\lambda = 1550$ nm. The sampled grating period $\Lambda_s$ calculated from Eq. (4) is 5 μm. Figure 3b depicts the configuration of the 4PS-TISE modulation. The initial sampled grating is a non-modulated 4PS design, comprising four π/2-phase (Φ) shift segments within $\Lambda_s$. The shift step $\Delta\Lambda$, determines the adjustment of the π/2-phase shift positions across each $\Lambda_s$. After each modulation period, $\Lambda_m$, $\Delta\Lambda$ resets to equal $\Lambda_s$, initiating a new modulation cycle. The total length of $L_M$ is equal to N × $\Lambda_m$. Here, we set $\Lambda_m = 20\Lambda_s = 100$ μm, $\Delta\Lambda = \Lambda_s/20 = 251$ nm. $L_M$ is changed from 0 μm (N = 0) to 400 μm (N = 4). Figure 3c illustrates the corresponding simulated photon density distributions over a 2 mm cavity length. It is noted that a longer TISE modulation 4PS grating region improves the field uniformity, meanwhile enabling SLM operation. Figure 3d shows the transmission spectrum with different $L_M$ and investigates the influence of the TISE on the SMSR. The highest SMSR occurs at 0.8 for $L_M = 200$ μm.

Figure 4a indicates the schematic structure of the 1D-TISE-PC laser based on the 4PS sampling structure. Figure 4b presents a scanning electron micrograph (SEM) side view image of the fabricated 1D-TISE-PC waveguide, demonstrating smooth and vertical sidewalls. This minimizes scattering loss thereby ensuring the designed performance is maintained. Figure 4c shows typical current-light (*I-P*) characteristics of the 1D-TISE-PC laser for different $L_M$ lengths. The threshold current was recorded at 50 mA, with the output power reaching 20 mW per facet at a DFB laser current ($I_{DFB}$) of 500 mA. The *I-P* curve exhibits fewer kinks and the highest slope efficiency with $L_M$ = 200 μm.

The 3D and 2D optical spectrum for the $L_M$ = 200 device is shown in Figs. 4d and 4e, with spectra measured at a resolution bandwidth (RBW) of 0.06 nm. Despite the non-coated laser facets, the results demonstrate lasing in a stable SLM with a high SMSR of 50 dB at $I_{DFB}$=200 mA, as shown in Figure 4f. Compared with other $L_M$ lengths presented in Fig. 4g-j. Mode hopping is reduced with increasing $L_M$, illustrating that the extension of TIS mode mitigates against depletion of the optical gain at the cavity center and suppresses competition from multi-longitudinal modes induced by SHB. It is also observed that with a larger $L_M$, the competition between the DFB and Fabry–Perot (FP) modes become more pronounced. This is attributed to the decrease in $\kappa L$ as $L_M$ increases in Eq. (3). Utilizing AR coatings would reduce facet reflection and eliminate the FP mode competition.

Linewidth broadening in semiconductor lasers is caused by non-uniformities in the intra-cavity photon distribution and SHB and follows the Schawlow-Townes linewidth strategy (see supplementary Part E). For further investigation of the suppression of SHB, we measured the linewidth of 1D-TISE-PC lasers using the fiber delayed self-heterodyne method (see supplementary Part F). The measured radio frequency (RF) spectra were fitted to Lorentz profiles as presented in Fig. 4k. The –3 dB linewidth was obtained by measuring the –10 dB bandwidth and dividing it by $\sqrt{9}$ to reduce the measurement error [30]. Figure 4l shows the −3-dB linewidth as a function of $I_{DFB}$ with different $L_M$. The linewidth reduces rapidly when $I_{DFB}$ is increased from 75 mA to 150 mA. Beyond this point, the value of the linewidth remains stable below 250 kHz from $I_{DFB}$ = 150 to 300 mA As the $I_{DFB}$ continues to increase, the linewidth increases due to the "linewidth floor" caused by SHB effects at high injection currents. It is found that inserting the TISE cavity effectively raises the upper current limit before the linewidth floor is reached, indicating that the TISE cavity suppresses SHB. However, the Fabry–Perot (FP) mode caused by an excessively long $L_M$ also broadens the linewidth at high injection currents. We also compared the 3$^{rd}$ order grating 1D-TISE-PC with a traditional 1$^{st}$ order non-TISE uniform grating laser with same epi structure and cavity length. Our results indicate a significant reduction in linewidth for the low coupling optical cavity design (see Supplementary Part E). To further improve the reliability of the results, we selected 15 device samples per $L_M$ value and measured their linewidths at $I_{DFB}$ = 150, 200, and 250 mA as shown in Fig. 4m. The narrowest average linewidth of 150 kHz was observed at $L_M$ = 200 μm; increasing or reducing $L_M$ resulted in linewidth broadening.

**Fig. 3: Design of the 1D-TISE-PC semiconductor laser structure based on 4PS sampling structure**

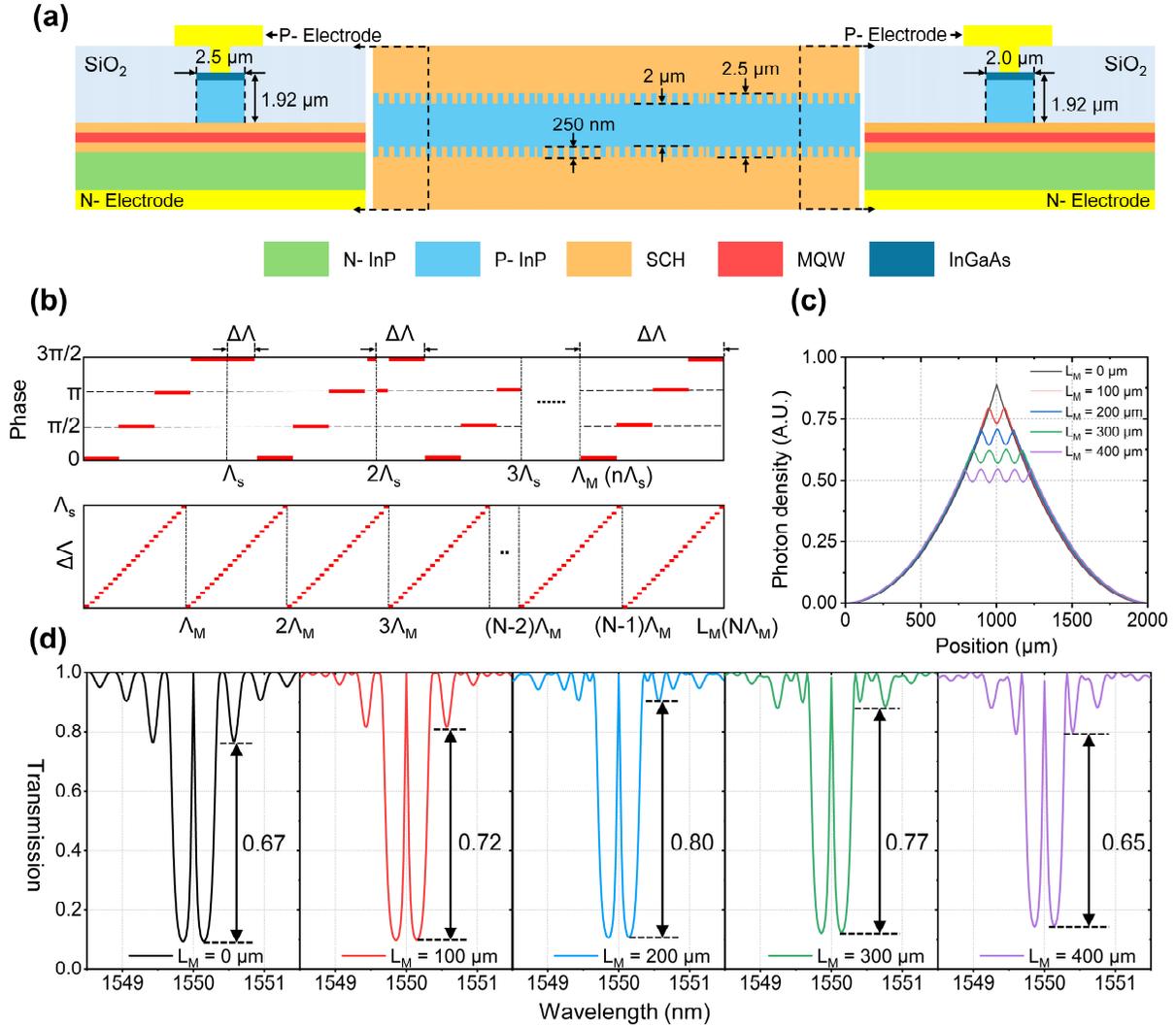

**a** The detailed device geometry and doping configuration of the 1D-TISE-PC semiconductor laser. **b** Modulation processing of the 1D-TISE-PC based on 4PS sampling grating structure. **c** Simulated photon density distributions and **d** transmission spectrum in the 2 mm length laser with $L_M$ changed from 0 to 400 μm.

**Fig. 4: Fabrication and static characterization of a 1D-TISE-PC semiconductor laser**

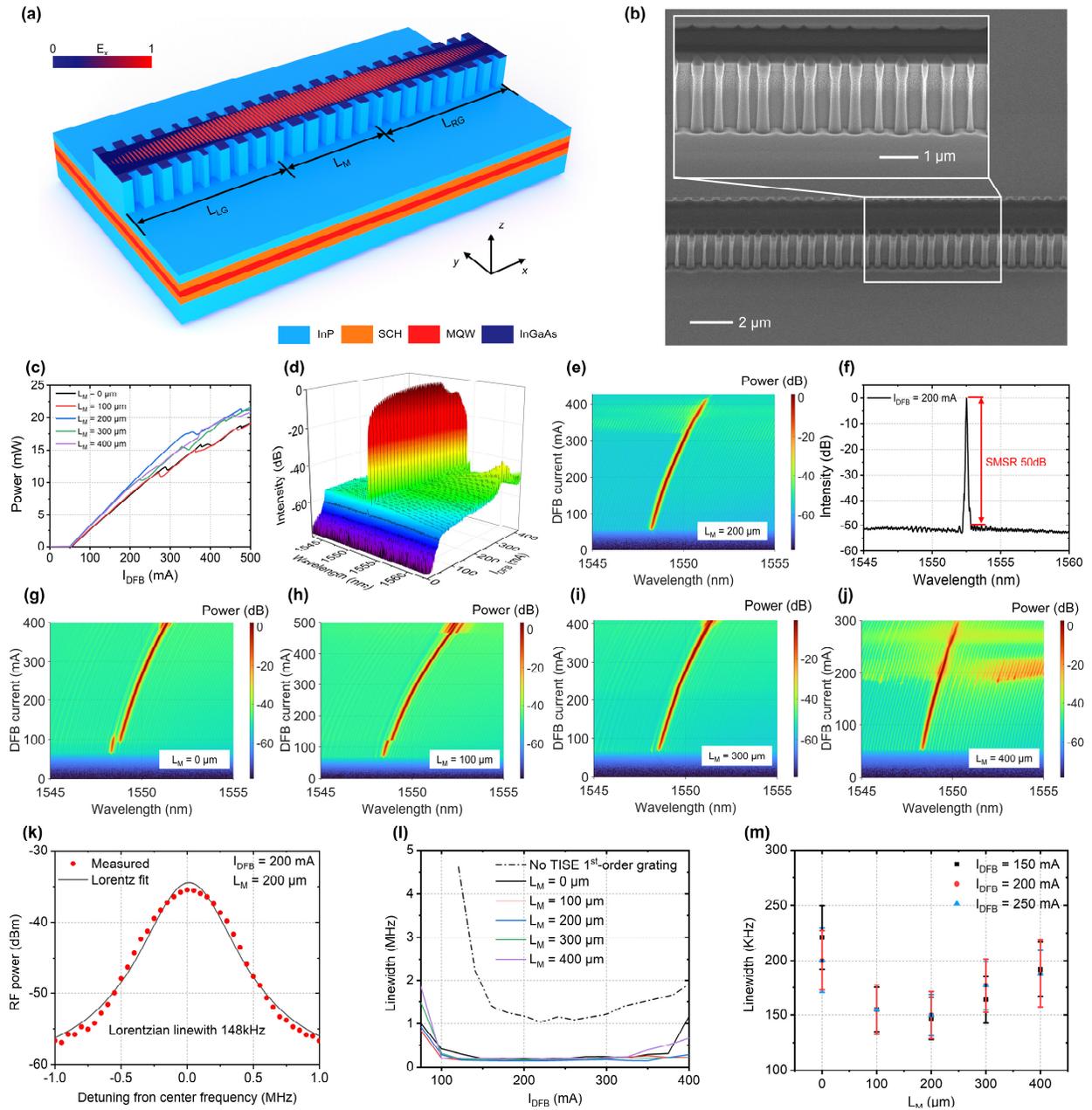

**a** Schematic of the 1D-TISE-PC laser based on a 4PS sampled structure. **b** Scanning electron micrograph (SEM) side view image of the fabricated 1D-TISE-PC waveguide. **c** Typical *I-P* characteristics with different $L_M$. **d** 3D optical and **e** 2D spectrum with $L_M$ = 200 μm. **f** Optical spectrum with $L_M$ = 200 μm and $I_{DFB}$ = 200 mA. Comparison of optical spectra with **g** $L_M$ = 0 μm, **h** $L_M$ = 100 μm, **i** $L_M$ = 300 μm, and **j** $L_M$

= 400 μm. **k** Measured Lorentz fitted linewidth with $L_M$ = 200 μm and $I_{DFB}$ = 200 mA. **l** Measured linewidth from $L_M$ = 0 μm to $L_M$ = 400 μm, compared with a 1$^{st}$ order grating laser with the same epi structure and cavity length and without a TISE cavity. **m** Measured linewidths for $I_{DFB}$ = 150 mA, 200mA and 250 mA with 15 device samples per $L_M$

## Multi-wavelength 1D-TISE-PC laser array

The 4PS 1D-TISE-PC structure can achieve precise control of the center wavelength through reconstruction equivalent-chirp (REC) technology. We designed an eight-channel 1D-TISE-PC laser array (Ch.1 to Ch.8) as shown in Fig. 5a. The sample grating period $\Lambda_s$ for Ch.1 to Ch.8 varied from 5 μm to 4.623 μm, with $L_M$ set at 200 μm. The time-delay spectra for different channels are shown in Fig. 5b, demonstrating that the variation in $\Lambda_s$ results in a 0.8 nm wavelength spacing. Figure 5c shows the optical spectra for each channel with $I_{DFB}$ = 200 mA. The corresponding linear fit to the lasing wavelengths is indicated in Fig. 5d where the slope of the line is 0.796 nm/channel with an error of 0.004 nm compared to the designed wavelength spacing of 0.8 nm. This error is less than the RBW (0.06 nm) of the optical spectrum analyzer (OSA), demonstrating excellent wavelength precision is achievable with the 4PS-SBG structure.

The SMSRs of the eight devices for $I_{DFB}$ = 100, 200, and 300 mA are shown in Fig. 5e. The SMSRs of all channels are around 50 dB. The linewidths of the fabricated 1D-TISE-PC laser array are shown in Fig. 5f. Each channel includes 10 device samples, and we observe a trend of increasing linewidth for channels with longer wavelengths. This is because the center wavelength is closer to the edge of the optical gain spectrum, where the linewidth enhancement factor, which is proportional to the derivative of the optical gain, increases and broadens the linewidth. Nevertheless, each channel has a linewidth lower than 300 kHz at $I_{DFB}$ = 200 mA.

**Fig. 5: Eight-channel 1D-TISE-PC laser array**

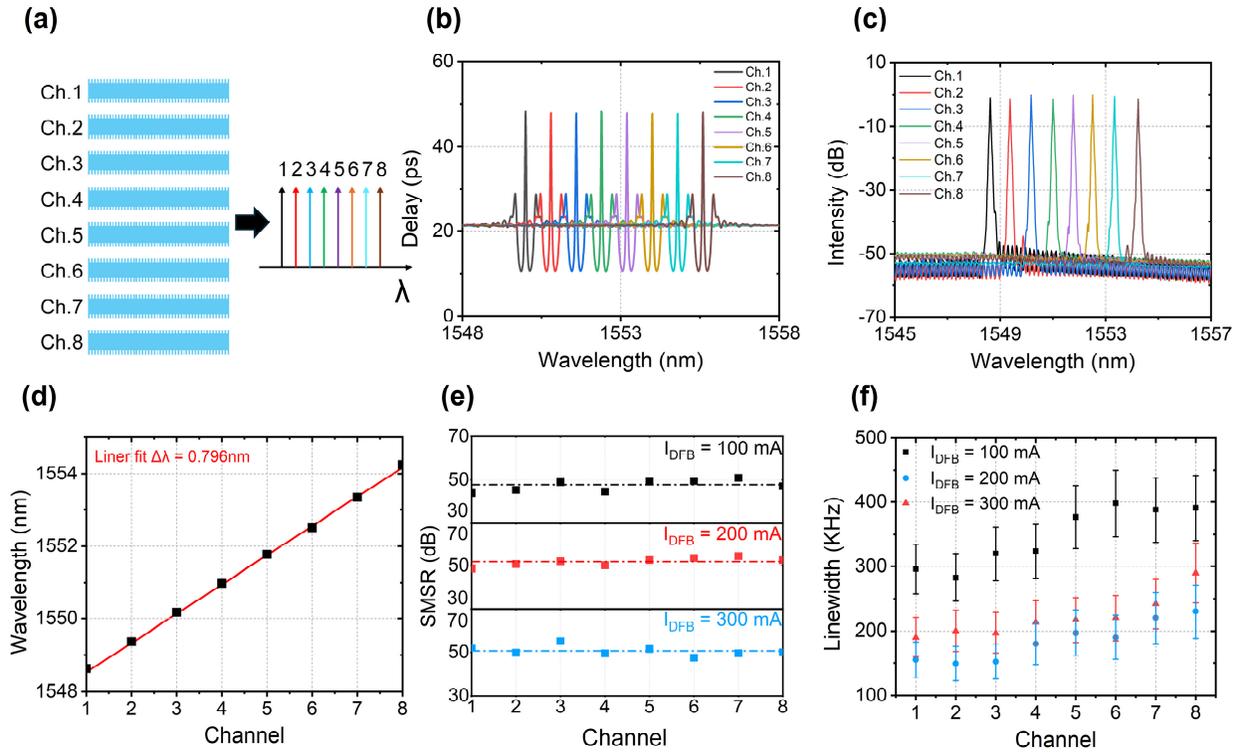

**a** The scheme of the eight-channel 1D-TISE-PC laser array d. **b** Time-delay spectrum for different channels. **c** Measured optical spectra of all the channels when $I_{DFB}$ = 200 mA **d** Lasing wavelengths and linear fit to the eight points. **e** SMSRs of eight lasers for $I_{DFB}$ = 100, 200, and 300 mA. **f** Measured linewidth of eight channels across a range of currents from 100 to 300 mA, with each channel comprising measurements from 10 device samples.

## Discussion

The concept of topology, which has recently been extended from solid-state materials to photonics, offers a powerful degree of freedom for manipulating light. It has been successfully employed to design robust optical cavities that form the foundation of topological micro- and nano-lasers operating in the infrared spectrum. In this work, we have implemented a 1D-TISE-PC grating structure for semiconductor lasers. This structure includes an energy band-degenerate TIS mode extension cavity, which efficiently extends the fundamental optical mode from the center cavity to the edges. As a result, the photon density along the cavity becomes flatter, reducing depletion of the optical gain in the central area.

Furthermore, we have combined the 1D-TISE-PC grating with a 4PS sampling grating structure to relax fabrication tolerances and achieve precise wavelength control. Specifically, we have experimentally demonstrated that the 1D-TISE-PC structure suppresses SHB at high current injection. The length-optimized TISE cavity can reduce the linewidth, minimize mode hopping, and apodize the side longitudinal mode. The 1D-TISE-PC laser exhibits SLM operation over a wide range of operating currents from 60 mA to 420 mA, even without AR facet coatings, achieving an SMSR of 50 dB and a narrow linewidth of 150 kHz.

Additionally, we have demonstrated an eight-channel 1D-TISE-PC laser array with a channel wavelength spacing of 0.796 nm, maintaining narrow linewidths and high SMSRs. This 1D-TISE-PC structure is compatible with commercial semiconductor laser fabrication procedures and can be extended to all topological one-dimensional lattice photonic crystal structures. Our work surpasses traditional designs using π-phase shift TIS modes and highlights the significant potential of the 1D-TISE-PC in advancing high-power and narrow-linewidth semiconductor laser development.

## Methods

### Device fabrication

The fabrication procedure for the 1D-TISE-PC laser can be found in[31]. The wafer was grown on an InP substrate using metalorganic vapor-phase epitaxy (MOVPE). The room temperature photoluminescence (PL) peak of the quantum wells (QWs) was located at a wavelength of 1530 nm. The 1D-TISE-PC sidewall grating waveguide was defined by electron-beam lithography (EBL) on an EBPG5200 E-beam system, with negative-tone Hydrogen Silsesquioxane (HSQ) acting as both the EBL resist and a hard mask for inductively coupled plasma (ICP) dry etching using a $Cl_2/CH_4/H_2$ gas mixture in an Oxford PlasmaPro 300 system. Subsequent steps included PECVD deposition of $SiO_2$ (Oxford PlasmaPro 100 PECVD), application of HSQ passivation layers, SiO$_2$ window opening, P-contact deposition, substrate thinning, and N-contact deposition, all performed using conventional laser diode (LD) fabrication techniques. Transmission electron microscopy images were acquired using a Hitachi SU8240 scanning electron microscope operating at 100 kV.

## Calculations and numerical simulations

The band diagrams of the topological photonic crystal were calculated by the coupling wave transfer matrix method (see Supplementary Part B). The spatial electric field distribution was obtained by using both the coupled wave transfer matrix method and the finite element method. The transmission spectra were calculated by the coupled wave transfer matrix method.

## Measurement setup

As shown in Supplementary Part F, the laser was driven by a continuous-wave current source (Newport Model 8000). An isolator was placed at the laser output. The light passed through the isolator and entered coupler 1, which had a splitting ratio of 50:50, dividing the light into two paths. One path included a 12.5 km long time-delay fiber and a variable optical attenuator (Thorlabs VOA50), while the other path included an 80-MHz acoustic-optic modulator (FIBER-Q T-M080) and a polarization controller (FPC032). These two paths were then combined in coupler 2 and split again into two paths: one leading to a photodetector (Thorlabs DET08CFC) and the other to an optical spectrum analyzer (Agilent 86140B). Both the current driver and the electronic spectrum analyzer (Keysight N9000B) were controlled by a computer via the general-purpose interface bus (GPIB) interface using LabVIEW software.

## **Data Availability**

The data that support the findings of this study are available from the corresponding author upon reasonable request.

## Acknowledgment

The authors would like to thank the staff of the James Watt Nanofabrication Centre at the University of Glasgow for their help in fabricating the devices, and the Critical Technologies Accelerator for support of this research.